Controllable and Fast Growth of High-Quality Atomically Thin and Atomically Flat Bi$_2$O$_2$Se Films

Yusen Feng,[1,‡] Pei Chen,[1,‡] Nian Li,[1,a)] Suzhe Liang,[2)] Ke Zhang,[1,a)] Minghui Xu,[1)] Yan Zhao,[1)] Jie Gong,[1)] Shu Zhang,[1)] Huaqian Leng,[1)] Yuanyuan Zhou,[1)] Yong Wang,[1)] and Liang Qiao[1,a)]

**Affiliations:**

[1)] School of Physics, University of Electronic Science and Technology of China, Chengdu 610054, China

[2)] Technical University of Munich, TUM School of Natural Sciences, Chair for Functional Materials, James-Franck-Str. 1 85748 Garching, Germany

[‡)] These authors contributed equally.

[a)] Authors to whom correspondence should be addressed: nianli@uestc.edu.cn (N. L.); phyzk@uestc.edu.cn (K. Z) and liang.qiao@uestc.edu.cn (L. Q.)

**Abstract**

As a novel and promising 2D material, bismuth oxyselenide (Bi$_2$O$_2$Se) has demonstrated significant potential to overcome existing technical barriers in various electronic device applications, due to its unique physical properties like high symmetry, adjustable electronic structure, ultra-high electron mobility. However, the rapid growth of Bi$_2$O$_2$Se films down to a few atomic layers with precise control remains a significant challenge. In this work, the growth of two-dimensional (2D) Bi$_2$O$_2$Se thin films by the pulsed laser deposition (PLD) method is systematically investigated. By controlling temperature, oxygen pressure, laser energy density and laser emission frequency, we successfully prepare atomically thin and flat Bi$_2$O$_2$Se (001) thin films on the (001) surface of SrTiO$_3$. Importantly, we provide a fundamental and unique perspective toward understanding the growth process of atomically thin and flat Bi$_2$O$_2$Se films, and the growth process can be primarily summarized into four steps: i) anisotropic non-spontaneous nucleation preferentially along the step roots; ii) monolayer Bi$_2$O$_2$Se nanosheets expanding across the surrounding area, and eventually covering the entire STO substrate step; iii) vertical growth of Bi$_2$O$_2$Se monolayer in a 2D Frank-van der Merwe (FM) epitaxial growth, and iv) with a layer-by-layer 2D FM growth mode, ultimately producing an atomically flat and epitaxially aligned thin film. Moreover, the combined results of the crystallinity quality, surface morphology and the chemical states manifest the successful PLD-growth of high-quality Bi$_2$O$_2$Se films in a controllable and fast mode.

**Introduction**

Due to the merits of moderate bandgap, excellent stability, and abundant nature retention, silicon has been a dominated semiconductor in microelectronic industry for more than 60 years.[1,2] However, when the thickness is less than 10 nm, silicon shows a reduced carrier mobility and a serious short-channel effect, which limits the further development of the silicon-based industry. In order to solve these problems and broaden Moore's Law, novel atomically thin 2D materials linked by weak van der Walls (vdW) force substitution of strong chemical bonding connections have received a lot of attentions, such as transition metal dichalcogenides ($MoS_2$ and $WS_2$), topological insulators ($Bi_2Te_3$ and $Bi_2Se_3$), black phosphorus and monolayer elemental films (Ge, Sn, silicene, etc).[3–6] Recently, a layered material, bismuth oxyselenide ($Bi_2O_2Se$) is regarded as one of the most promising candidates for the next-generation electronic industry due to its high electron mobility (450 cm$^2$ V$^{-1}$ s$^{-1}$, room temperature),[7,8] moderate and tunable bandgap (~0.8 eV, up to 1.5 eV depending on thickness),[3,7,9] and excellent ambient stability.[7,10,11] Besides, $Bi_2O_2Se$ can also be used in many other fields, including memristors,[12,13] ultrafast spectroscopy,[14] photocatalysis,[15] terahertz switches,[16] field-effect transistors,[7] photodetector,[17,18] solar cells,[19] etc. The prerequisite for these various applications is the fabrication of $Bi_2O_2Se$ materials. Although most study focus on the bulk crystals and multiplayer films, achievement of atomically thin $Bi_2O_2Se$ films and atomically flat $Bi_2O_2Se$ films deserves more efforts, with consideration of a large quantity of use in electronic industry.

Generally, $Bi_2O_2Se$ films can be achieved via various methods, such as chemical vapor deposition (CVD),[7] molecular beam epitaxy (MBE),[20] pulsed laser deposition (PLD),[10,21] and solution-assisted methods. However, only the self-limiting vapor-solid deposition and MBE methods have been successfully applied for the atomical growth of low-dimensional $Bi_2O_2Se$ thin films. For example, Khan et al. used the self-limiting vapor-solid method to obtain few-layer single crystal domains on a mica substrate;[22] Liang et al. exploited MBE to precisely control the growth of monolayer $Bi_2O_2Se$ on a SrTiO$_3$ (STO) substrate.[20] Compared to these two methods (CVD and MBE), PLD can combine advantages of the fast and controllable growth. However, there have been no reported on PLD-grown $Bi_2O_2Se$ film with a monolayer or atomically flat surface of a large-scale area and a high quality. Importantly, the growth process for the $Bi_2O_2Se$ films down to a few atomic layers has not been revealed in detail so far.

Therefore, in this work, we present a PLD approach to grow atomically thin and flat $Bi_2O_2Se$ films with high quality in a large-scale area. We thoroughly investigate the effects of various parameters on the regulation of film growth by the PLD method. With precise optimization of the parameters, including growth temperature, laser energy density, laser emission frequency and oxygen pressure, a high-quality growth mode of 2D atomically flat $Bi_2O_2Se$ thin films is achieved on a (001)-oriented STO substrate in a fast speed. Furthermore, the key growth processes of $Bi_2O_2Se$ on STO are elucidated with the aid of atomic force microscopy (AFM), X-ray diffraction (XRD), and high-resolution X-ray photoelectron spectroscopy (HRXPS), which also provide the evidence of the quality of obtained atomically thin and atomically flat films. Notably, the atomically flat film structure is clearly and directly visible in the AFM images. An intertwined weave structure is observed on the Bi$_2$O$_2$Se films prepared by the PLD method, which, in conjunction with other results, demonstrates the successful fabrication of higher-quality Bi$_2$O$_2$Se films using PLD in a controllable and fast mode, as compared to MBE and CVD. This knowledge is expected as an essential piece required to broaden the application of Bi$_2$O$_2$Se films.

## Results and Discussion

As illustrated in Figure 1a, the as-obtained target is mounted to the rotator and subjected to laser illumination with a stable emission frequency and emission intensity provided by 248 nm wavelength KrF laser. High vacuum level of 10$^{-4}$ Pa is reached inside the entire chamber and after that oxygen is introduced, and target-substrate distance is set to 5.0 cm. Upon laser irradiation, the high-energetic species ablated from the target condense on the etched STO substrate, and then Bi$_2$O$_2$Se/STO films can be obtained after cooling. By rationally adjusting the parameters of temperature, oxygen pressure, energy density and emission frequency, we can obtain a series of films. Figure 1b shows the crystal structures of Bi$_2$O$_2$Se and STO. As a cubic perovskite oxide, STO (001) has a lattice parameter of a = 3.90 Å, which provides perfect lattice matching with Bi$_2$O$_2$Se ($\approx$ 0.5% mismatch) of a tetragonal structure (I4/mmm, a = b = 3.887 Å, c = 12.202 Å, Z = 2).[23] Scanning transmission electron microscopy (STEM) and First-Principle calculation provide an insight that Bi-TiO$_2$ is the most closely bound way of interface epitaxial growth for Bi$_2$O$_2$Se.[20,24]

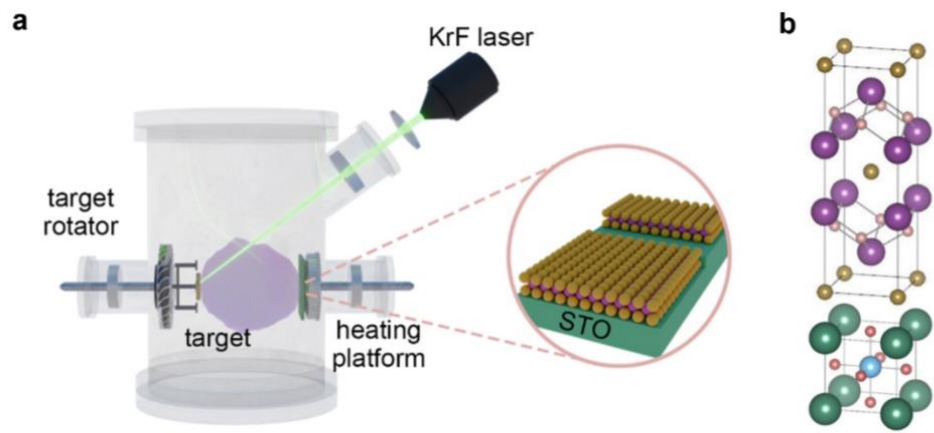

**Figure 1.** a) Schematic illustration of epitaxial growth of atomically flat Bi$_2$O$_2$Se on the (001)-oriented STO substrate by PLD method, b) crystal structure of Bi$_2$O$_2$Se and STO.

Upon reaching the substrate, the species or atoms would undergo various processes, including surface diffusion, surface migration, re-evaporation, adsorption, and condensation.[25] During epitaxial growth, nucleation accompanied by condensation is considered a non-spontaneous process.[26] From a thermodynamic perspective, by analyzing the Gibbs free energy change and the energy equilibrium condition at all interfaces, we can determine that a lower temperature reduces the critical size required for nucleation, resulting in a lower critical free energy barrier. This, in turn, benefits for the formation of thin, continuous films with fine grains. However, if the temperature is too low, it can hinder surface diffusion, leading to condensation at low-energy chemisorption sites, which disrupts epitaxial growth and crystallization. If temperatures closer to the equilibrium temperature can produce coarser single-crystal structures. But, excessively high temperatures can exacerbate re-evaporation, and thus it is challenging to accomplish adsorption and condensation processes.[25] As shown in Figure 2a, the XRD patterns of $Bi_2O_2Se$ films prepared at different temperatures indicate that single crystals can be formed at 400-510 °C, while it is difficult to condense into a film at 550 °C. We can also know from Figure 2a that the intensity of diffraction peak and its full with at half maximum (FWHM) are optimized at about 450 °C. Figure 2b shows the amplification of the (004) diffraction peaks at temperatures of 430, 450, 470, and 490 °C. The clearly visible Laue oscillations and Kiessig fringes suggest that a high and coherent out-of-plane order over the entire thickness.[10] The emergence of Laue oscillations at the sample of 450 °C is generally taken as the evidence of the homogeneity of the film and smoothness of the interface, thus implying that 450 °C is an optimal growth temperature (Figure 2b). In addition, the AFM images of the films at different temperatures (Fig. S1) can support that 450 °C plays as a critical temperature for the surface roughness of the films. That is a local minimum root mean square (RMS) value at 450 °C.

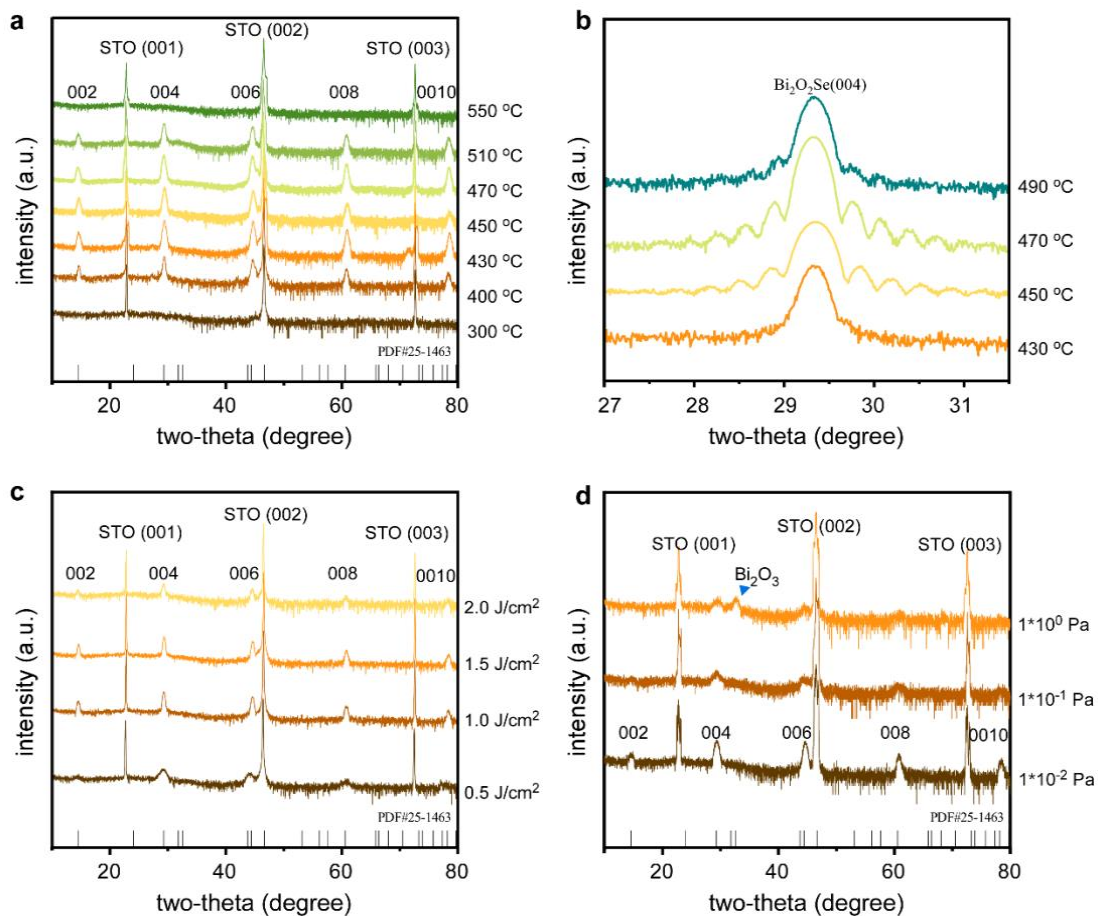

**Figure 2.** Effects of different conditions on the growth of thin films. a) Out-of-plane XRD patterns of $Bi_2O_2Se$ films prepared at different temperatures. b) $Bi_2O_2Se$ (004) XRD peak of films grown at different temperatures of 430 °C, 450 °C, 470 °C, 490 °C. Laue oscillations indicate a high and coherent out-of-plane order over the entire thickness. c) XRD patterns of $Bi_2O_2Se$ films prepared with different laser energy densities. d) XRD patterns of $Bi_2O_2Se$ films prepared at oxygen pressures of $1\times10^0$ Pa, $1\times10^{-1}$ Pa, and $1\times10^{-2}$ Pa.

In contrast to temperature, the deposition rate exhibits a negative correlation with both the critical nucleus size and the change in the free energy of nucleation. Therefore, to achieve fine-grained continuous films, a higher deposition rate is required, which can be facilitated by employing lasers with high emission frequency and high energy density, as well as reducing the oxygen pressure.[25] Through $Bi_2O_2Se$ (006) diffraction peaks, we obtain lattice constants and grain sizes at different emission frequencies by Bragg's Law ($2d\sin\theta = n\lambda$) and Scherrer Formula ($D = \frac{k\lambda}{\beta cos\theta}$). Our research investigates a range of laser emission frequencies from 1 Hz to 9 Hz, and laser energy densities from 0.5 J/cm² to 2.0 J/cm². XRD data collected under all these experimental conditions (Fig. S2a and Figure 2c) consistently show the formation of single-crystal structures. Fig. S2b,c display that with emission frequency increase, the crystallite and growth rate increase as well, but the lattice constant show a fluctuation trend. Thus, the high emission frequency of 9 Hz can improve the crystallinity of the $Bi_2O_2Se$ film, and simultaneously we can obtain a relatively low RMS roughness value (Fig. S3). Similarly, as the laser energy density increases (Fig. S4), the growth rate overall increases but the lattice constant decreases and gradually gets closer to the theoretical value (12.164 Å). However, as shown in Fig. S5, a high laser energy density adopted here is unsuitable, which shows a high RMS value. We assume that a high laser energy density will promote surface diffusion and grain coalescence (known as the Osvaldo annexation or fusion). This process contains the merging of smaller grains to form larger grains, thus further disrupting surface flatness and the 2D Frank-van der Merwe (FM) growth mode, and finally leading to an island growth.

Tuning the oxygen pressure to affect surface diffusion in a way that promotes the desired crystal structure and smooth film growth. Figure 2d and Fig. S6 show the XRD and AFM of the films prepared at oxygen pressures of $1 \times 10^0$ Pa, $1 \times 10^{-1}$ Pa and $1 \times 10^{-2}$ Pa, respectively. The AFM results indicate 3D Volmer Weber (VW) growth mode at $1 \times 10^{-2}$ Pa oxygen pressures, at which $Bi_2O_2Se$ is deposited in the form of island-like particles, and that may be related to the vacancy of oxygen atoms due to the insufficient oxygen involved in the reaction. Whereas, 2D FM growth mode is exhibited at $1 \times 10^{-1}$ Pa oxygen pressure (Fig. S6b,e; terrace-like structure observed), which is more favorable for us to grow high-quality atomically thin and atomically flat films. $1 \times 10^0$ Pa oxygen pressure will also cause growth mode of 3D VW, due to the fact that excess oxygen leads to the formation of heterogeneous phases. Besides the surface morphologies, we continue to compare the crystal phase among these three samples. Besides the $Bi_2O_2Se$ phase, we can also find $Bi_2O_3$ diffraction peaks in the $1 \times 10^0$ Pa sample, which can prove that the impurity phase is formed (denoted with an arrow in Figure 2d).

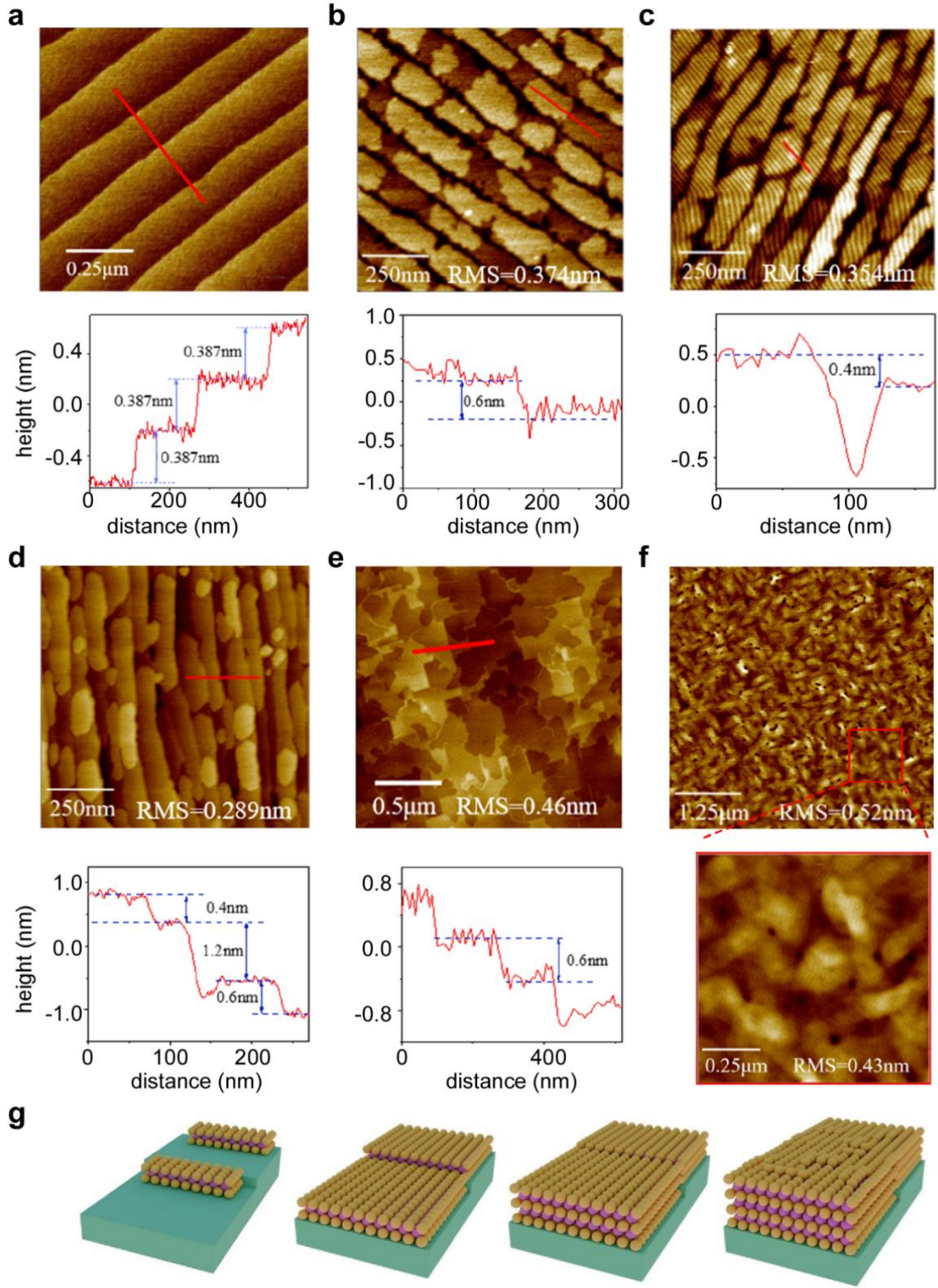

**Figure 3.** Growth processes of atomically flat $Bi_2O_2Se$. a) Surface morphology of $TiO_2$-terminated STO substrate, showing clear terraces with a step height of 0.387 nm. b-f) Surface morphology of $Bi_2O_2Se$ films with different thicknesses: b,c) monolayer, d) 6 nm, e) 147 and f) 35 nm. g) Schematic illustration of $Bi_2O_2Se$ film grown on the (001) surface of STO.

Following a comprehensive investigation of each individual parameter, including single-crystal growth rate, grain size, flatness, and epitaxial growth quality, we have identified an optimal growth environment: a temperature of 450 ℃, an oxygen pressure of $3.7 \times 10^{-1}$ Pa, and a laser energy density of 0.5 J/cm$^2$ with an emission frequency of 9 Hz. This condition is employed to grow a series of $Bi_2O_2Se$ thin films with varying thicknesses on etched STO (001) substrates with stepped $TiO_2$ termination. Previous studies have reported that, $TiO_2$-terminated STO readily coordinates with $Bi_2O_2^{2+}$ to form a compound $[BiTiO_4]^{1-}$,[27] which facilitates the binding of $Bi_2O_2Se$ to STO substrate. Furthermore, the creation of a 2D electron gas (2DEG) requires an STO substrate with $TiO_2$ termination,[28] which exhibits an exceptionally high carrier mobility (exceeding $10^4$ cm$^2$ V$^{-1}$ s$^{-1}$) at the perovskite heterojunction.[28–30] To prepare etched substrates, we use a BHF buffer solution etching method (more details provided in Supplementary Information, Experiment section). As displayed in Figure 3a, high-quality etched substrates with a step width around 150 nm are chosen. The AFM pattern shows that the chosen substrates have an excellent flatness with a step height of 0.387 nm, which is comparable to the lattice constant of perovskite-structure STO lattice and also to the spacing of the two $TiO_2$ termination.

Figures 3b-d and 3g illustrate the AFM images and a schematic diagram of the growth process of $Bi_2O_2Se$ thin films. During the initial nucleation stage, $Bi_2O_2Se$ exhibits preferential growth along the step roots, forming monolayer $Bi_2O_2Se$ nanosheets with a rim. This process creates a new step in the original plane, with a height of approximately 0.6 nm (Figure 3b), corresponding to the thickness of a combined $[Bi_2O_2]^{2n+}$ and $[Se]^{2n-}$ layer. Subsequently, the single-layer $Bi_2O_2Se$ nanosheets will expand to the surrounding area. As they grow, adjacent nanosheets will merge and eventually cover the entire STO step. At this stage, a V-shaped trench forms, comprising the $Bi_2O_2Se$ step, the STO step, and the $Bi_2O_2Se$ step on the next layer. Notably, the height difference between the two $Bi_2O_2Se$ steps is consistent with the STO layer height difference of 0.387 nm.

As the sputtered $Bi_2O_2Se$ plasma continues to be deposited, the monolayer grows upward in a 2D FM mode. Although ions in the V-trench have the ability to form bonds on multiple sides, the 0.389 nm height difference between adjacent $SrTiO_3$ steps causes the upper and lower $Bi_2O_2Se$ layers to become misaligned. Furthermore, the identical height of the upper $[Bi_2O_2]^{2n+}$ layer and lower $[Se]^{2n-}$ layer would result in a significant increase in interfacial energy if newly adsorbed atoms were to condense into the V-shaped trench, thereby increasing the total energy of the system. As a result, the V-shaped trench is retained. As shown in Figure 3c, lattice mismatch leads to gaps in the same layer of $Bi_2O_2Se$, and then the film will continue to generate multiple layers upwards. Figure 3d shows that the preferential growth of some regions creates a larger step height difference, which would make it possible to grow along the step gradient. If the transverse growth exceeds the width of the next step, it can connect to the step further down, forming a transverse sheet. Ultimately, the diffusion and connection of the sheets along the step with the transverse sheet give rise to a crossed textile structure. The AFM pattern of textile structure can be seen in Figure 3f. According to previous studies,[31] Se vacancies can also form textile structures, but it is not clear whether there is any correlation between the two textile structures. It can be seen that 147 nm-thick Bi$_2$O$_2$Se films do not show strip-like steps, but exhibits a regional lamellar structure (Figure 3e). The height of adjacent steps is still consistent with the height of single Bi$_2$O$_2$Se layer, which suggests the film well maintains a 2D FM layer-by-layer growth mode and that the mismatches between the regions have been resolved. X-ray reflectivity (XRR) measurements (Fig. S7) reveal the presence of oscillations in the reflectivity curve for the 147 nm thick film, which signifies a coherent out-of-plane structural order throughout the entire thickness of the film. All these samples show a single crystal phase without impurities (Fig. S8).

To have an in-depth understanding about the crystallinity of the Bi$_2$O$_2$Se samples, the rocking curves of the Bi$_2$O$_2$Se (004) peaks are examined. The ω-rocking curve of the Bi$_2$O$_2$Se (004) peak, as shown in Figure 4a, is relatively narrow, and has a FWHM of ~0.07°, consistent with the fact that Bi$_2$O$_2$Se deposited on STO has a low lattice mismatch and grow structurally similar directions. We further examined the in-plane crystal structure of Bi$_2$O$_2$Se epitaxial thin films deposited at 450 °C. As shown in azimuth Φ scans of the off-axis {011} peaks of STO substrate and the Bi$_2$O$_2$Se film (Figure 4b), the four peaks separated by 90° are present in these two films, which confirms the in-plane alignment of the Bi$_2$O$_2$Se film with the STO substrate.[10,20] The Bi$_2$O$_2$Se and STO peaks are located at the same angular position, indicating that the unit cells show no rotation with respect to one another. Thus, we can conclude that the Bi$_2$O$_2$Se films deposited on STO via PLD method show a high crystal quality. Figure 4c shows the growth rate of Bi$_2$O$_2$Se film grown on STO is about 5.9 nm/min, which is about 11 times faster than that of multilayer Bi$_2$O$_2$Se/STO films realized by the CVD method and about 196 times faster than the MBE method.[20,21] Higher growth rates mean lower production costs, which suggests that synthesis of high-quality Bi$_2$O$_2$Se films by PLD method holds excellent promise for industrial applications.

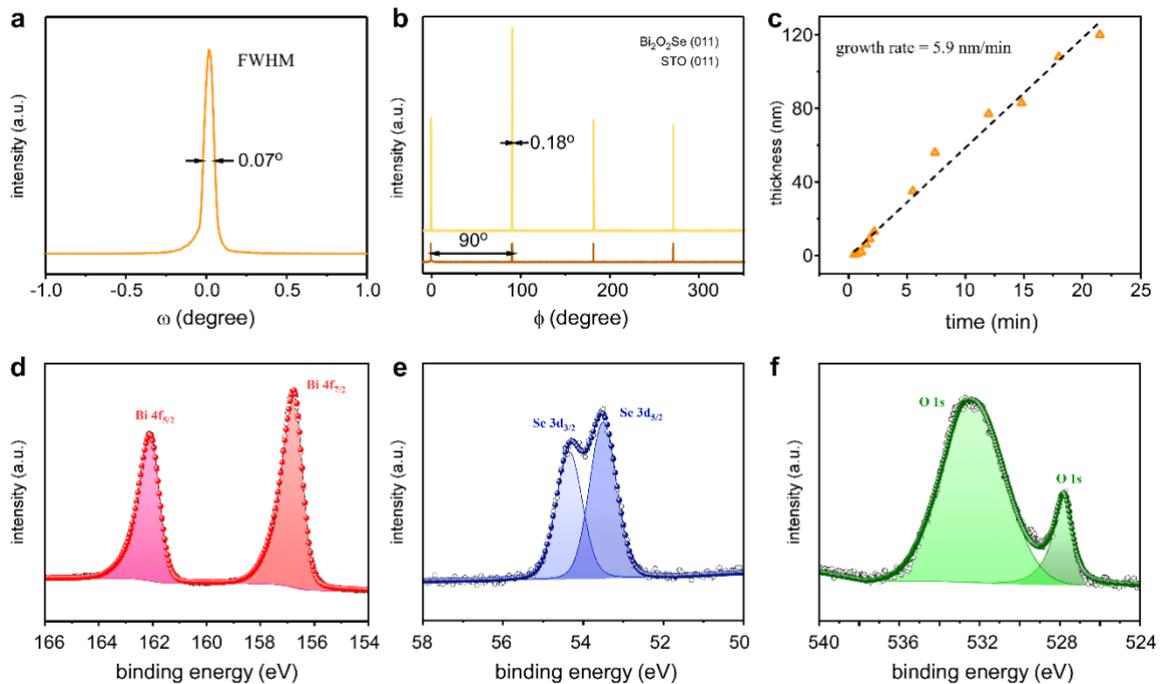

**Figure 4.** Quality inspection of the films grown at 450 °C. a) The ω–rocking curve of corresponding (004) reflection. b) Azimuth Φ scans of the off-axis {011} peaks of STO substrate and the $Bi_2O_2Se$ film. c) Plot of film thickness versus deposition time, showing a growth rate of 5.9 nm/min. d-f) X-ray photoemission spectra of 35 nm-thick $Bi_2O_2Se$/STO film.

The electronic properties of 35 nm $Bi_2O_2Se$ film are characterized using HRXPS. As shown in Figure 4d, the two spin-orbital-coupling (SOC) symmetric peaks of Bi $4f_{7/2}$ and Bi $4f_{5/2}$ are centered at 156.8 eV and 162.1 eV, which can be attributed to the Bi peak in the O-Bi bond.[3,32,33] Similarly, according to the effect of SOC, the Se 3d peak is split into two Lorentz peaks centered at 53.5 eV and 54.4 eV (Figure 4e), corresponding to the Se $3d_{5/2}$ and Se $3d_{3/2}$.[3,32] The fitted results of the peaks are all within a reasonable chemical shift range,[34,35] which is also confirmed by the HRXPS data of the $Bi_2O_2Se$ film at different specific thicknesses (Fig. S9). Thus, we reason that the $Bi_2O_2Se$ film have a high quality in its chemical state.

### Conclusion

In conclusion, we have systematically explored the impact of various fabrication parameters, involving growth temperature, laser energy density, laser emission frequency and oxygen pressure, on the film properties by PLD method and using etched STO (001) substrates, and simultaneously have gained a deep understanding of the growth mechanism for $Bi_2O_2Se$ films. The growth process can be primarily summarized into four steps: 1) anisotropic non-spontaneous nucleation preferentially along the step roots; 2) monolayer $Bi_2O_2Se$ nanosheets expanding across the surrounding area, and eventually cover the entire STO substrate step; 3) monolayer $Bi_2O_2Se$ grows vertically in a 2D FM epitaxial growth, and 4) with a layer-by-layer 2D FM growth mode, ultimately producing an atomically flat and epitaxially aligned thin film. By precisely controlling the thickness of $Bi_2O_2Se$ films grown via a 2D FM mode, we successfully prepared atomically thin and atomically flat $Bi_2O_2Se$ films on STO substrate with $TiO_2$ termination, exhibiting epitaxial quality superior to that of previously reported $Bi_2O_2Se$/STO films grown by PLD methods. Notably, the film growth rate in the present study reaches 5.9 nm/min, offering a significant cost advantage over other methods, like MBE, for large-scale industrial fabrication. Furthermore, XRD and HRXPS measurements verify that the $Bi_2O_2Se$ films exhibit excellent epitaxial growth with high crystallinity quality and chemical homogeneity.

### Supplementary Material

See supplementary material for the experimental details, the surface morphology of $Bi_2O_2Se$ films at different temperatures, XRD data of $Bi_2O_2Se$ films at different emission frequencies, XRD data of $Bi_2O_2Se$ films at different laser energy densities, surface morphology of $Bi_2O_2Se$ films at different laser energy densities, surface morphology of $Bi_2O_2Se$ films at different oxygen pressures, X-ray reflected pattern of different thicknesses and XPS data of $Bi_2O_2Se$ films at representative thicknesses.

## Acknowledgements

This work was supported by funding from the National Science Fundation of China (12304078, 52072059)，Natural Science Foundation of Sichuan Province (2024NSFSC1384).

## AUTHOR DECLARATIONS

### Conflict of Interest

The authors have no conflicts to disclose.

### Author Contributions

Yusen Feng and Pei Chen equally contributed to this work.

**Yusen Feng:** Data curation (supporting); Formal analysis (equal); Investigation (equal); Methodology (equal); Writing -original draft (main); Writing – review & editing (equal). **Pei Chen:** Data curation (main); Formal analysis (equal); Investigation (equal); Methodology (equal). **Nian Li:** Conceptualization (equal); Data curation (supporting); Investigation (equal); Methodology (equal); Project administration (equal); Supervision (equal); Writing -original draft (supporting); Writing – review & editing (equal). **Suzhe Liang:** Data curation (supporting); Visualization (equal); Writing – review & editing (equal). **Ke Zhang:** Conceptualization (supporting); Investigation (equal); Methodology (equal); Project administration (equal); Supervision (equal); Writing -original draft (supporting); Writing – review & editing (equal). **Yan Zhao:** Investigation (equal); Writing – review & editing (supporting). **Minghui Xu:** Investigation (equal); Writing – review & editing (supporting). **Jie Gong:** Investigation (equal); Writing – review & editing (supporting). **Shu Zhang:** Investigation (equal); Writing – review & editing (supporting). **Huaqian Leng:** Formal analysis (equal); Writing – review & editing (supporting). **Yuanyuan Zhou:** Formal analysis (equal); Writing – review & editing (supporting). **Yong Wang:** Formal analysis (equal); Writing – review & editing (supporting). **Liang Qiao:** Conceptualization (equal); Data curation (supporting); Formal analysis (equal); Funding acquisition (main); Investigation (equal); Methodology (equal); Project administration (equal); Resources (main); Supervision (equal); Validation (equal); Writing -original draft (supporting); Writing – review & editing (equal).

## DATA AVAILABILITY

The data that support the findings of this study are available from the corresponding author upon reasonable request.